\documentclass[sigconf]{acmart}
\renewcommand\footnotetextcopyrightpermission[1]{} 
\usepackage[normalem]{ulem}
\usepackage{xspace}
\usepackage{lipsum}
\usepackage{multirow}
\usepackage{colortbl}
\usepackage{tabularx}
\usepackage{colortbl}
\usepackage{subcaption}
\usepackage{booktabs}
\usepackage{graphicx}
\usepackage[usestackEOL]{stackengine}
\usepackage{stfloats}
\usepackage{balance}
\usepackage{url}
\usepackage{fixltx2e}
\usepackage{tikz}
\usepackage{pgfmath}
\usetikzlibrary{decorations.text, arrows.meta,calc,shadows.blur,shadings}

\usepackage{shortcuts}

\begin{document}

\title{Quantifying Carbon Emissions due to Online Third-Party Tracking}

\author{Michalis Pachilakis, Savino Dambra\\
Iskander Sanchez-Rola, Leyla Bilge\\
\textit{Norton Research Group}}

 \begin{abstract} 
	In the past decade, global warming made several headlines and turned the
	attention of the whole world to it.  Carbon footprint is the main factor
	that drives greenhouse emissions up and results in the temperature increase
	of the planet with dire consequences.  While the attention of the public
	is turned to reducing carbon emissions by transportation, food consumption
	and household activities, we ignore the contribution of \COtwo emissions produced
	by online activities.  In the current information era, we spend a big amount
	of our days browsing online. This activity consumes electricity which in
	turn produces \COtwo.  Using the Internet is something that we cannot avoid
	but several of the things
	happening behind the scenes during browsing are further contributing to \COtwo
	emissions production.  While website browsing contributes to the production
	of greenhouse gas emissions, the impact of the Internet on the environment
	is further exacerbated by the web-tracking practice.  Indeed, most webpages are
	heavily loaded
	by tracking content used mostly for advertising, data analytics and usability
	improvements.  This extra content implies big data transmissions which
	results in higher electricity consumption and thus higher greenhouse gas
	emissions. 
	
	In this work, we focus on the overhead caused by the web-tracking practice and
	analyze both its network and carbon footprint. By leveraging the browsing
	telemetry of \mbox{100k} users and the results of a crawling experiment of
	\mbox{2.7M} websites, we find that web tracking increases data transmissions
	upwards of 21\%, which in turn implies the additional emission of around 11 Mt of greenhouse gases
	in the atmosphere every year. We find such contribution to be far from
	negligible, and comparable to many activities of modern life, 
	such as meat production, transportation, and even cryptocurrency mining.
	Our study also highlights that there exist significant inequalities when
	considering the footprint of different countries, website categories, 
	and tracking organizations, with a few actors contributing to a much greater
	extent than the remaining ones.
\end{abstract}
\maketitle
\pagestyle{plain}

\section{Introduction} \label{sec:intro}

Global warming has constantly been in the spotlight during the last
decade~\cite{globalwarming}. Countless articles have discussed its main
consequences, from the reduction of glaciers to the extinction of animal
species, and from more intense heat waves to the sea-level rise. 
Combating climate change has become a priority, and many countries try to put
in place incentives aimed at reducing their polluting
emissions~\cite{globalwarminggovern}. Although this requires a collective
effort on both national and international sides, it substantially boils down to
reducing the carbon footprint of each individual.

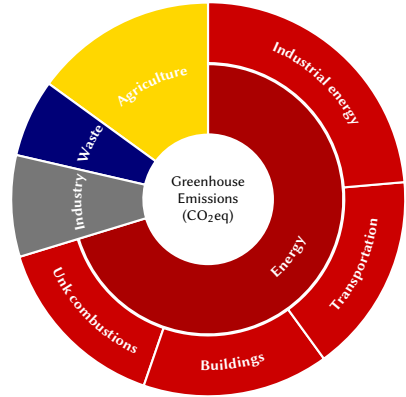
\begin{figure} \centering 
    \resizebox{0.25\textheight}{!}{
    \begin{tikzpicture}[
		font=\sf    \scriptsize,
		myarrow/.style={ thick, -latex, }, Center/.style ={ circle, fill=white,
		text=root, align=center, font =\footnotesize, inner sep=1pt,          },
		scale=1.1   
	]

	\node[Center](ROOT) at (0,0) {Greenhouse \\ Emissions \\ (\COtwo)};


	\arctext[N1][red1][15pt](1.5)(-163)(90){|\footnotesize\bf\color{white}|
	Energy};
		\arctext[N2S1][red2][13pt](2.5)(-163)(-109){|\footnotesize\bf\color{white}|
		Unk combustions};
		\arctext[N2S2][red2][13pt](2.5)(-109)(-54){|\footnotesize\bf\color{white}|Buildings};
		\arctext[N2S3][red2][13pt](2.5)(-54)(5){|\footnotesize\bf\color{white}|
		Transportation};
		\arctext[N2S4][red2][13pt](2.5)(90)(5){|\footnotesize\bf\color{white}|
		Industrial energy};
	\arctext[N2][gold1][28.2pt](1.96)(144)(90){|\footnotesize\bf\color{white}|
	Agriculture};
	\arctext[N3][blue1][28.2pt](1.96)(167)(144){|\footnotesize\bf\color{white}|
	Waste};

	\arctext[N4][gray1][28.2pt](1.96)(197)(167){|\footnotesize\bf\color{white}|
	Industry};

	\end{tikzpicture}
}
	\caption{Carbon footprint breakdown. Energy represents the major factor with
	72.3\% of the emissions, followed by Agriculture (18.4\%), Industry (\COtwo
	emitted by chemical conversions - 5.2\%) and Waste (methane and nitrous
	oxide emitted due to decomposition of organic residues - 3.2\%).}
	\label{fig:carbonFootprint} \end{figure}

The carbon footprint of an individual is calculated by estimating the total
amount of greenhouse gases (\COtwo\footnote{CO\textsubscript{2} equivalent is a
	metric measure used to compare the emissions from various greenhouse gases
	(\GHG) on the basis of their global-warming potential}) that are produced due to
necessities of being alive and of being an active member of the current
society~\cite{naturecfp}. Globally, those emissions are estimated to be 
over 34 billion tonnes per year~\cite{owidco2andothergreenhousegasemissions}.
Energy production and consumption represent the major factors that contribute to
carbon emissions, followed by those caused by
agricultural processing, industrial conversions, and waste decomposition
(Figure~\ref{fig:carbonFootprint}).

With the emergence of computers and mobile devices, individuals started to spend
a significant amount of their time using the Internet~\cite{internetspendtime}.
Thanks to the Internet, we are able to complete everyday errands from the comfort
of our house, such as buying groceries and attending meetings with someone on the other
side of the world.  While seemingly the usage of the Internet helps to reduce
\COtwo emissions, people might easily overlook that online activities
require electricity and consequently generate \COtwo as well.  Over the last two
decades, we observed a dramatic increase in Internet
usage~\cite{activeInternet}, and experienced another exaggerated rise during the
COVID-19 pandemic. According to a recent report~\cite{internetusage}, the
average user spends 7 hours per day connected to the Internet.
While website browsing contributes to the production of
\GHG, the impact of the Internet on the environment is further exacerbated by
the web-tracking practice. Indeed, the large majority of web pages are heavily
loaded by third-party trackers~\cite{1m_measurement,sanchez2018knockin}
that constantly track users for various reasons, including
advertisement, analytics, and usability
improvements~\cite{lerner2016internet}. The retrieval and execution of
tracking-related content implies data transmissions and the use of computing
resources, thus resulting in higher electricity consumption and \GHG emissions. 

Measuring the power consumption of the Internet and the corresponding \COtwo
emissions are two overly complex problems.
A large corpus of studies has already tried to calculate the power consumption
in order to transfer 1GB of data on the Internet. But due to different
methodologies and system boundaries in each study (e.g., data centers, underwater
cables, ISPs) their estimations may vary even up to 5 orders of
magnitude. In addition, estimating the \COtwo emissions of the Internet is a
much more understudied area that requires an accurate conversion from energy
consumption to \GHG emission. This needs to take into account a plethora of
factors, the most important being the mixture of sources used to produce
electricity in each country.

The most recent reports consider the Internet responsible for 3.7\% of the
global \GHG emissions ($\sim$1 gigatonne of \COtwo)~\cite{co2internet}. More
specifically, only two studies narrowed their estimation down to web tracking.
P\"{a}rssinen et al. provided an upper bound estimation of 159
Mt~\cite{parssinen2018environmental} for the \COtwo emissions caused by the
online advertisement ecosystem, indicating that almost 16\% of the global
Internet's electricity consumption is due to the third-party trackers. 
However, authors only considered a worldwide average for quantifying the amount
of \COtwo emitted for each kWh; on top of that, they in turn relied on previous
estimations for converting the amount of transmitted data to the energy
consumed, which introduces a non-negligible uncertainty in their results (i.e.
from 20 to 282 TWh consumed for advertisements only).
The findings of Cucchietti et al. revealed a more conservative
estimation~\cite{cucchietti2022tracking}. The authors conducted an experimental
study to measure the monthly network traffic produced by third-party cookies on
the Tranco top 1M websites and estimated that 11.4 kt (kilotonnes) of \COtwo per
month is caused by the transfer of cookies. 
However, considering the uncertainty ranges from the sources used, they estimate
a much larger spectrum (between  1.4 kt to 17.1 kt of \COtwo per month).

Since this study only accounts for the size of cookies, the overall consumption is
significantly lower than the previous study~\cite{parssinen2018environmental}
---even though the authors use a much higher estimation of the kWh consumed per GB of
data transferred.
The two studies differ by several orders of magnitude in their final findings since they measure
different aspects of the tracking ecosystem, use estimations to understand the 
user behavior and lack real-world data.
On top of that, both studies suffer from big flunctuations in their final findings as lower and upper bound
for \COtwo emissions, which comes as a result of the estimations and global averages used during the calculations.

To overcome the limitations that characterize those works,
we designed an experimental study that analyzed real-browsing
telemetry of almost 100k users to perform more accurate estimations of both
network and \COtwo footprints due to third-party tracking.
At first, the telemetry made it possible to capture real-users browsing behaviors without
the need for approximations. Differently from previous studies, we also accounted for
potentially-cached content, investigated tracking-related headers, and
differentiated results by website categories and tracker organizations.
In addition, we paid close attention to the geographical dimension in our
measurements, as computing web-tracking footprint is a multidimensional problem that
cannot be addressed by blindly considering global averages. In this respect, 
we looked at the overload that trackers introduce in different
geographical locations, and estimated \GHG emissions by considering the energy production
sources of the country and continent each user mainly browses from.

Below we report our most significant findings on the web-tracking footprint: 
\begin{itemize} 
	\item The overall emissions due to trackers for the global active Internet 
	users account for 10.79 Mt of \GHG (6.28 Mt for the most conservative computation).  
	\item The average annual data transmitted for a single user on the Internet
	is 8.14 GB, out of which 1.67 GB for tracking-related purposes only (21.11\%). 
	\item Asian users emit more \COtwo compared to users from other continents mostly because of the resources used in energy production. 
	\item Shopping webpages are characterized by the highest mean data
	transmitted for a single user due to tracking (182 MB yearly). 
	Websites serving News and media content reveal the highest average ratio 
	of tracking-related content per website (2.2 MB per page).
	When looking at emissions, Shopping websites represent the biggest
	offender with 0.82 Mt \COtwo every year
	\item Google emissions, the largest tracker, annually account for 4.38 Mt \COtwo
	(40\% of the global emissions for web tracking)
	\item Globally, web tracking annually generates \GHG as 80\% of the whole
	aviation system of the top-10 polluting countries for plane emissions.
	Similarly, its impact is comparable to 70\% of the emissions produced by
	Bitcoin mining in the US. Electricity required by the tracking ecosystem
	could heat New York for half a year.
\end{itemize}
The remaining of the paper is structured as follows.  In
Section~\ref{sec:carbonfp}, we list the polluting impact that human's essential
activities have on the environment. We then describe the main datasets used in
this study together with methodology details in Section~\ref{sec:method}.  We
report the results of our experiment in (Section~\ref{sec:results}). In
Section~\ref{sec:comparisons}, we compare web-tracking environmental impact
with those characterizing human kind's principal necessities. In
~\ref{sec:related} we report the related work, discuss the main implications
and limitations of our work in Section~\ref{sec:discussion}, and conclude in
Section~\ref{sec:conclusions}.

\section{The Carbon Footprint of a Human Life} 
\label{sec:carbonfp} 
In this section, we report figures on essential activities that characterize
individuals of the modern society, such as food production, power/electricity
consumption and the use of different transportation means. We then focus on the
cyberspace and report the consequences of the massive adoption of the Internet
and its facilities on the global gas emissions.  We analyze previous surveys
and provide an equation that quantifies the energy consumption ---and in turn
the \COtwo emissions--- for each GB of data transferred on the network.  In
addition, we investigate the environmental impact of Bitcoin adoption and
report the carbon footprint resulting from its mining activity.  We finally
focus on some of the natural carbon dioxide absorbents by describing their
contribution to capturing or converting greenhouse gases from the atmosphere. 
In the following paragraphs, we discuss per-capita \COtwo emissions for the
countries that contribute the most to global pollution, and report data about
the biggest offenders. After presenting web-tracking footprint in
section~\ref{sec:results}, we will then provide emission comparisons in
Section~\ref{sec:comparisons}.

\subsection{Per-capita \COtwo emissions}
According to the Emissions Database for
Global Atmospheric Research (EDGAR), each person in the world produces 4.79
tons of \COtwo every year~\cite{edgarReport}, 
with an annual increase of 0.9\% from 2020 to 2021.
While this represents the average global value, there exist significant
inequalities across countries, mainly due to the different living standards and
sources used to produce energy. For example, while the carbon footprint for a
Qatar citizen reaches upwards of 37 tonnes of \COtwo per year, an inhabitant of
Mali is responsible for the emission of 0.09 tons in the same
period~\cite{worldmeter}. Globally, energy production (electricity for
buildings, transportation, and industrial applications) is the main polluting
factor and it is responsible for 73.2\% of the greenhouse gas emissions.
On the other hand, activities related to
agriculture, direct industrial processes, and waste contribute respectively
18.4\%, 5.2\% and 3.2\% of the total emission.

\begin{table*}
	\centering
	\caption{Estimated emissions (tons \COtwo per-capita) for human essential activities.}
	\label{tab:co2_consumptions_and_emissions}
		\begin{tabular}{lcccccccccc}
			\toprule
			\textbf{Emissions}			&  \multicolumn{10}{c}{\textbf{Country}}														\\
			\textbf{source} 			&  China &     US &  India &  Russia &  Japan & Germany &  Canada &   Iran &  S. Korea &  Indonesia \\
			\midrule
			Any 			 &  7.38 &  15.52 &  1.91 &  11.44 &   9.70 &     9.44 & \underline{18.58} &  8.08 &    11.85 &      2.03 \\
			Meat             &  2.90 &   \underline{5.80} &  0.20 &   3.50 &   2.40 &     4.80 &    4.10 &  1.90 &     3.40 &      0.80 \\
			Passenger car    &  0.50 &   \underline{4.49} &  0.20 &   1.00 &   1.40 &     1.80 &    4.10 &  1.60 &     2.00 &      0.50 \\
			Aviation         &  0.09 &   0.58 &  0.02 &   0.15 &   0.14 &     \underline{0.70} &    0.50 &  0.03 &     0.40 &      0.04 \\
			Electricity      &  2.70 &   \underline{5.40} &  0.63 &   2.10 &   3.50 &     2.40 &    1.90 &  1.80 &     4.70 &      0.70 \\
			\bottomrule
		\end{tabular}
\end{table*}

\subsection{Foodprint: Carbon Footprint of What We Eat} 
Food production accounts
for 25\% of the world's \GHG emissions and currently requires half of the
earth's surface~\cite{poore2018reducing}. 
The whole process has a substantial environmental cost,
quantified at 13.7 billion tons of \COtwo produced every
year~\cite{owidenvironmentalimpactsoffood}. \GHG emissions due to food
production are mainly caused by land use and processes at the farm stage, such
as the application of fertilizers and enteric fermentation ---i.e., methane
produced in the stomachs of the cattle.  The combination of land use and
farm-stage emissions account for upwards of 80\% of the
foodprints~\cite{owidenvironmentalimpactsoffood}. On the contrary, food
transportation and supply chain-related activities are small contributors (10\%
of the emissions each).

The carbon footprint varies a lot among food types: while meat, cheese, and eggs
have the highest impact, fruit, vegetables, beansi, and nuts produce much lower
\GHG. Meat production unquestionably stands out as the most polluting activity,
as animals live longer than plants, imply the destruction of forests to make way
for pasture and produce as a result of the digestive process large quantities of
methane ---which is 34 times more polluting than
CO\textsubscript{2}~\cite{gotBeef}. According to a study published by Poor et
al.~\cite{poore2018reducing}, beef meat is by far the largest offender,
generating 60 kg/\COtwo per kilogram of meat produced which results in more than
double the emissions of the second biggest polluter, lamb meat (23 kg/\COtwo per
kilogram of product).  Plant-based foods have a significantly lower impact,
causing on average the production of 1 kg/\COtwo per kg of product. Annually,
approximately 100 megatonnes of \COtwo are emitted due to the production of
beans and nuts, which result in 0.33\% of the total \COtwo emission versus 15\%
that is caused by meat production and consumption~\cite{poore2018reducing}.

Apart from the food type, emissions also depend on the dietary habits of people
with different nationalities, ideologies, or religions. For example, the United
States and Australia are the countries with the highest average meat
consumption ---and consequent \COtwo emitted--- per capita, 
with 98.21 kg/year and 94.04 kg/year respectively. This is more than double the
global average ---41.90 kg/year per capita--- and more than 26 times compared to
India, where the consumption per capita is limited to 3.63 kg/year.

\subsection{Transportation carbon footprint}
\begin{table}[!htbp]
	\centering
	\caption{Share and \COtwo emissions of each transportation sector}
	\label{tab:co2_transportation}
		\begin{tabular}{lcc}
			\toprule
			\multirow{3}{*}{Sector} 		&   Share in the 		& \COtwo  				\\
			&   emissions 			& emissions 			\\
			&   (\%)				& (tons)	\\
			\midrule
			Passenger cars 					&   41\%  				&  2.99 B 	\\
			Medium/heavy trucks				& 	22\%  				&  1.60 B 	\\
			Shipping 						&   11\% 				&  0.80 B  	\\
			Aviation 						&   8\% 				&  0.58 B   \\
			Buses and minibuses				&   7\% 				&  0.51 B	\\
			Light vehicles					&   5\% 				&  0.36 B 	\\
			2/3 wheelers					&   3\% 				&  0.22 B 	\\
			Rail							&   3\% 				&  0.22 B 	\\
			\bottomrule
		\end{tabular}
\end{table}
In 2020, transportation has
been responsible for the production of approximately 7.3 billion metric tons of
CO\textsubscript{2}~\cite{transportBreakdown}. As reported in
Table~\ref{tab:co2_transportation} of the Appendix, passenger cars are responsible for 41\% of
the emissions, followed by medium and heavy trucks (22\%) and cargo carriers
(11\%). Aviation and rail had a lighter impact on the environment, producing 8\%
and 3\% of the polluting gases.

\texttt{Car transportation} - In 2019, the European Environment Agency (EEA)
quantified in 122.3 g \COtwo/km the average emissions of passenger
cars~\cite{transportBreakdown}. Similar to the foodprints, vehicle emissions per
capita vary across the globe (Table~\ref{tab:co2_consumptions_and_emissions}). 
The US have the highest emissions per-capita, with the average citizen emitting 4,486 kg of \COtwo annually,
followed by Canada (4,120 kg \COtwo) and Saudi Arabia (3,961 kg \COtwo). \GHG
produced are considerably lower in low-income countries such as those in South
Asia and Africa, where the average person produces approximately 70 times less
emissions annually compared US (e.g., 21 kg \COtwo in Congo and 48 kg \COtwo in
Eritrea).

\texttt{Air transportation} - According to the International Council of Clean
Transportation, domestic and international flights globally account for around
2.5\% of \COtwo emissions in the atmosphere every year~\cite{flyingFootprint}.
The estimation of \GHG produced by a plane to cover one kilometer is slightly
lower than the one computed for an average passenger car: a Boeing 737-400
releases 115 g/km \COtwo per passenger (90 kg \COtwo per hour) considering a
cruising speed of 780 km/hour~\cite{boingConsumption}. The global average of
emissions due to aviation is around 103 kg \COtwo per capita each
year~\cite{flyingFootprint}. However, some countries are responsible for a much
larger share of emissions.  In the United Arab Emirates, each individual is
responsible for the production of 1,950 kg of \COtwo per year, almost 20 times
more than the global average. On the other hand, in countries such as India,
Morocco and Croatia the average person is responsible for only 17 kg of \COtwo.

\subsection{Electricity carbon footprint} 
In a survey conducted by the Energy
Information Administration (EIA), the worldwide cumulative electricity
consumption reported during 2019 is 23.4 PWh~\cite{electricityConsumption}.
China is the highest consumer (8.3 PWh) followed by the US (3.9 PWh). But even so, this
does not correctly reflect the per-capita consumption, as most of the energy
consumed in China is used for industrial activities. In this respect, Iceland,
Norway, and Bahrain represent the countries where the average citizen consumes
the most electricity, respectively 52 MWh, 23 MWh, and 17 MWh --- with China
being only 57th in the ranking of per-capita
consumption~\cite{electricityPerCapita}.  The lowest electricity consumption is
registered in countries with lower incomes and living standards, such as those
in central Africa (0.05 MWh per capita) and central Asia (0.15 MWh per capita).

The \COtwo emissions due to electricity generation strictly
depend on the sources and technologies used to produce the
energy~\cite{electricityFootprint}. Iceland is one of the greenest producers of
electricity thanks to its geothermal and hydro sources, only generating 28 g
\COtwo per kWh. The carbon footprint of Singapore is instead close to 497 g
\COtwo per kWh due to the use of coal and gas. For
these reasons, \GHG emitted in Iceland due the whole process reached 0.52 million
tons of \COtwo, whereas they account for more than double in the case of
Singapore (1.32 million tons of \COtwo).

\subsection{Carbon Footprint of Internet} 
Active Internet users reached 4.95
billion individuals worldwide in 2022~\cite{activeInternet}. Although web
facilities allow us to send messages, download music, stream videos, and share
pictures, they have a substantial impact on the environment. In 2020, network
infrastructures, data centers, large servers, and personal devices have been
found to be responsible for 3.7\% of global \GHG,  quantified in 1.6 billion
tons of \COtwo~\cite{internetImpact} . While the production of electricity for
keeping up and running all the devices has the biggest share in the pollution
equation, their manufacturing, shipping, and cooling play an important role as
well.  Despite the huge increase in data demand in the last decade --- 3 times
more from 2015 to 2020 --- \COtwo emissions due to internet traffic have
remained stable thanks to the increased efficiency of hardware
components~\cite{dataCenterImpact}. Nevertheless, this equilibrium is expected
to end at the end of this decade.

Worldwide, Asia and Pacific regions host the highest number of data centers (95)
followed by the United States and Canada (75)~\cite{dataCenterImpact}. In a
similar way, China and India in the Eastern Asia count the highest number of
active Internet users (respectively 1.01B and 0.83B
individuals)~\cite{countryInternetUsers}.

\subsection{Cryptocurrency carbon footprint}

With the increasing demand for cryptocurrencies such as
Bitcoin, Ethereum, and Monero, and the growing adoption of blockchain
technologies in several applications, we witnessed a dramatic increase in
coin-mining activities and transaction validations which were proven to consume
a considerable amount of electricity~\cite{DEVRIES2022498}. Cryptocurrencies
---in particular those based on a proof-of-work consensus mechanism that
requires the solution of complex mathematical puzzles--- cause a non-negligible
share of the polluting emissions.

Estimating the environmental impact of cryptocurrencies depends
on miners' geographical locations, and on the means that are used to produce
electricity in those countries (which might change from one region to the other
or over time).  As a matter of fact,
different studies report completely different statistics for Bitcoin's share in
the global \COtwo emissions~\cite{STOLL20191647,LI2019160,DEVRIES2022498}. 
According to one of the most recent and comprehensive study, almost 3M devices performed
crypto-mining in 2021 only, requiring 13 GW of
electricity~\cite{DEVRIES2022498}.
Because of the increased market demand and the miners' preference for cheaper but
less eco-friendly produced electricity, the average carbon intensity of
electricity consumed increased from 478.27 g\COtwo/kWh on average in 2020 to
557.76 g\COtwo/kWh in 2021. According to this estimation, 65.4 Mt of \COtwo
per year are produced due to Bitcoin activity. The big players that
contribute the most to the overall \COtwo emission are Kazakhstan (25.2), US
(15.1), Russia (8.9), Malaysia (3.4), Germany (2.0), Iran (1.9), Canada (1.5),
and Ireland (1.3). The rest of the countries contribute a combined 6.3 Mt
\COtwo per year due to Bitcoin activities.

\subsection{CO\textsubscript{2} natural absorbents}
Plants, oceans, and soil
represent the main natural carbon sinks whose combination is able to absorb
about half of all the carbon dioxide emissions from human
activities~\cite{friedlingstein2020global}.  While those absorbents constitute
valuable allies in fighting global pollution, their efficacy is threatened by
deforestation and by the saturation of soils and oceans due to the
ever-increasing quantities of \GHG in the atmosphere.

\texttt{Plants} - The world's forests absorb a total of 15.6 gigatonnes of
\COtwo per year (i.e., about 34\% of the total emissions in the same period) as
plants capture and use carbon dioxide during photosynthesis to produce
glucose~\cite{carbonSinks}. On average, a tree is able to absorb between 18 and
20 kg of \COtwo every year~\cite{treesAbsorb,kiran2011carbon}. 

\texttt{Soil} - Although soil does not absorb CO\textsubscript{2} as quickly as
vegetation, its storing capability is much higher: according to a study by Ontl
et. al., global soil contains almost twice as much CO\textsubscript{2} as the
living flora and the atmosphere combined~\cite{ontl2012soil}.

\texttt{Oceans} - Seas could theoretically absorb 95\% of human-made
\GHG~\cite{carbonSinks}. In reality, each year oceans are able to store a share
of 6.3\% of the total emissions (2.9 Gt of \COtwo per year)~\cite{carbonSinks},
because of the slowness of processes that transfer gases from the atmosphere to
water and deposit them on the seabed.

\section{Data Sources and Methodology} \label{sec:method}

This section summarizes the methodology and the datasets that constitute the
basis of our study. Our main dataset is the telemetry of
web-browsing logs provided by a popular antivirus company (Source A). 
To preserve the privacy of the users, only the domain name of the whole URL was
recorded excluding any Personal Identifiable Information (PII).  For each domain
in the telemetry, its corresponding category is also provided. We
visit all of the unique domains in our data through a custom crawler (Source B)
and capture all the HTTP requests performed to load the website and its
resources. We calculate the amount of network traffic produced by the trackers
identified in the crawled websites. Leveraging public reports and services to
assess the electricity cost across countries and the power
consumption due to the internet traffic (Source C), we estimate electricity
consumed due to the traffic produced by the trackers.  Finally, we convert the
consumed electricity to \COtwo emissions to understand how much tracking
activities contribute to the annual production.

\subsection{Web-browsing telemetry}
\label{sec:method-sub:telemetry} 

The telemetry at our disposal (Source A)
contains the web-browsing history of 100k users, respectively collected on 50k
desktop hosts and 50k mobile devices, for a period of 4 weeks (28 days) across
the year.  These logs are collected by the company's product installed on its
user devices.
The data is collected from users who voluntarily install the product, accept the
company's privacy policy~\cite{nortonPolicy}, and opt-in to share their data.
The customer's identifier is anonymized on the client-side and transmitted in
this form to a central data lake: in our analysis, we observe users only through
numeric anonymized identifiers that do not contain any attribute or detail that
allows us to trace back to their identities. 

In our study, we estimate the amount of traffic produced by the trackers in an
average week, and then calculate the amount for the whole year. To better
capture users' browsing behaviors along the year and mitigate potential trends
of accessing particular categories of websites tight to specific seasons or
events (e.g., summer or Christmas), we analyze 4 different weeks evenly
distributed from July 2021 to February 2022 (i.e., 7 consecutive days in 4
different months). Each entry in the telemetry consists of a code reporting the
country of the user, the unique user identifier, domain name, its category, and a timestamp.
Overall, the 4 weeks of data included 41M entries of 2.75M distinct domains
(3.44\% were not active anymore) that were later crawled to measure the produced
network traffic. We filtered out users whose browsing session was not able to be
reconstructed through our crawler (because all their accessed domains were not
reachable). Our final dataset consists of $48,931$ mobile and $47,995$ desktop
users.

\subsection{Reconstructing the Browsing Sessions}
\label{sec:method-sub:crawler} 

The most precise network traffic
production estimation can be done through live data collection at the time of
the browsing by the real users.  Unfortunately, this method is extremely
privacy-concerning and intrusive to the users, which could result in potential
data leaks and breach of data confidentiality.
For this, we try to reconstruct the data by crawling the domains with a custom
crawler (source B).
We crawled the domains in the first week of March 2022 using  a server powered
by two Intel Xeon E7-8890 v3 processors and 2Tb of RAM. We use a single machine
located in the US, as the variability of trackers encountered by scanning websites from
different locations has been proven to be negligible on large-scale
measurements~\cite{whenSally}.  

The crawler uses Puppeteer~\cite{puppeteer} together with a fresh instance of
Google Chrome for each website. All the requests and responses performed are
gathered by using the HTTP Archive (HAR) facility of the browser and stored in
a JSON format.  In order to avoid possible detections of our automated browser,
we instrument Puppeteer with a plugin that implements state-of-the-art
anti-detection techniques~\cite{puppeteerStealth}.
As we already mention, maintaining user privacy is our top concern and for this reason
we exclude the path after the domain name. 
This means that our crawler crawls the main page of a website, or a subdomain if this is part of the URL.
In that sense we manage to capture a lower bound of trackers, since some trackers (e.g. multimedia related) might be 
embedded deeper in the websites, or some websites might require login before accessing the content. 
For each domain, we analyze
the requests made to load third-party resources and identify those that belong
to known trackers~\cite{mozillaTracking,easyprivacy,sanchez2021journey}. 
Besides providing content for tracking purposes
(e.g. javascript libraries), trackers could also provide other functionalities to a website, such as
delivering images or videos. In our study, we excluded media content,
such as videos or images, since our goal is to only measure the impact of web tracking. 
Therefore, media content transmitted by the trackers and their impact on the network
traffic is out of scope for our paper. 
For example, \textit{trackerA} could deliver a video to be played in a publishers webpage.
This video is not directly related to tracking even though it is created by a
third-party tracker. It is rather needed for the functionality of the visited
website. Here, we do not record the video content related traffic. 
However, the headers that notify the tracker about the user's visit are accounted.
Since a single organization might own multiple domains, we also group trackers under 
the organizations they belong to by using the relationships
provided by three previous works.~\cite{mozillaTracking,easyprivacy,sanchez2021journey}. 

Additionally, in order to analyze the users' online behavior in the most
pragmatic manner, we stored the \texttt{Cache-Control} HTTP headers of all the
corresponding tracking domains flagged~\cite{cacheControl}. This information
will allow us to retroactively enforce those caching policies, and precisely
estimate the size associated with each user session in our dataset. More
concretely, we follow the multi-keyed HTTP cache model (also known as state or
cache partitioning~\cite{cachePartitioning}) recently implemented in all popular
browser families (i.e., Chrome, Firefox, and
Safari~\cite{chromeCache,firefoxCache,safariCache}). According to it, browsers cache 
content not only by taking into account the content itself, but also
where it is being loaded from (based on the eTLD+1 scheme~\cite{eTLD}). For example,
if both \textit{example.com} and \textit{test.com} load
\textit{content.com/script.js}, the resource will not be shared but cached individually. 
Furthermore, while some requests indicate long cache time periods (e.g.,
\texttt{max-age=31536000}), the cache is usually invalidated before by the
browsers, and content is stored not more than 47 hours on average~\cite{cache47}.
For this work, we consider this same upper bound limit to make results as
accurate as possible.

\subsection{Internet browsing emissions}
\label{sec:method-sub:electricity}

We calculate the \COtwo emission due to the
trackers on web browsing sessions by converting each GB transmitted over the
Internet to its corresponding electricity usage (source C). Among the studies published
since 2000, there is a non-negligible variance about the electricity-consumption
estimates of Internet traffic. In some cases, the difference is even up to 5
orders of magnitude, from 136kWh/GB to 0.004kWh/GB. This
dramatic variance is due to the differences in the system boundary and increased
efficiency over the last two decades~\cite{aslan2018electricity}. While some
studies only take into account the electricity consumed during the network
transmission, others also incorporate the electricity consumed by the data
centers, undersea cables, network accesses, ISPs, the type of device used, and
any other equipment involved. GB to kWh conversion
carries a long number of other limitations, such as the fact that it does not
take into account the distance between the source and the destination, and the fact
that the processing requirements of the data centers might not be linearly
correlated with the size of data.

The most recent studies that considered combinations of measurement approaches
and system boundaries reported the total electricity consumption of a 1 GB data 
transmission to be in the range of 2.48 and 7.1 
kWh~\cite{malmodin2014,costenaro2012,krug2014,malmodin2016}. 
If the electricity consumed by data centers, servers, 
computers, and ISPs are not incorporated, they reported 0.023 to 0.16 kWh/GB.
The most recent academic study analyzed data from 2015, therefore likely very
different estimations would be made with more recent data.  At a more recent
conference organized by the American Council on the subject of Energy-Efficient
Economy, on the other hand, it was reported a slightly lower number: 3.1
kWh/GB~\cite{standfordmag}.  Calculating the carbon footprint of the web has been an
interest to industrial and non-profit organizations as
well~\cite{websiteCarbon,wholegrain}. To calculate the carbon footprint of a
given website, companies use the 1.8 kWh/GB estimation based on the
predictions made about the electricity usage of communication technologies for
the next ten years~\cite{globalelectricityusage}. In our study, we will rely on
these two most recent estimations (1.8 kWh/GB, 3.1 kWh/GB) rather than the higher
ones reported with data from 2015. As deeply discussed by Aslan
et. al~\cite{aslan2018electricity}, the time factor greatly impacts the
estimations made by the previous works, and probably the current electricity
consumption per GB of data transmission is lower than the one of 7 years ago.

\begin{table}[!htbp]
	\centering
	\setlength{\tabcolsep}{1.8pt}
	\caption{Overview of continents and top-3 countries ordered by percentage of users.}
	\label{tab:results-countries}
		\begin{tabular}{lrlrr}
			\toprule
			Continent &   \% Users & Country &  \% Users & CO\textsubscript{2}/kWh\\
			\midrule
			&			&      United States 	&  11.7 \% & 0.45 kg\\
			North America 		&   24.0 \% &      Canada 			&   6.6 \% & 0.13 kg\\
			&			&      Mexico 			&   2.6 \% & 0.45 kg\\
			\midrule
			&			&      Great Britain 	&   1.4 \% & 0.35 kg\\
			Europe 				&  21.5 \% 	&      Germany 			&   1.2 \% & 0.38 kg\\
			&			&      Netherlands 		&   1.2 \% & 0.45 kg\\
			\midrule
			&			&      Japan 			&   9.0 \% & 0.5 kg\\
			Asia 				&  20.4 \% 	&      India 			&   1.9 \% & 0.7 kg\\
			&			&      Hong Kong 		&   1.7 \% & 0.81 kg\\
			\midrule
			&			&      Australia 		&  10.3 \% & 0.79 kg\\
			Oceania 			&  19.0 \% 	&      New Zealand 		&   5.7 \% & 0.09 kg\\
			&			&      Fiji 			&   0.1 \% & NA\\
			\midrule
			&			&      Brasil 			&   8.2 \% &  0.07 kg\\
			South America 		&  13.9 \% 	&      Chile 			&   0.9 \% & NA\\
			&			&      Colombia 		&   0.9 \% & NA\\
			\midrule
			&			&      South Africa 	&   0.8 \% & NA\\
			Africa 				&   1.3 \% 	&      Nigeria 			&   0.1 \% & NA\\
			&			&      Egypt 			&   0.1 \% & NA\\
			\bottomrule
		\end{tabular}
\end{table}
Once we convert GB to kWh, we have to estimate the \COtwo emissions, which could
significantly differ by region. In
Table~\ref{tab:results-countries}, we provide the amounts of \COtwo
per kWh produced for the top countries in our dataset.
Moreover, we found that the worldwide average of \COtwo emitted in order to
produce 1 kWh is 420 grams.
\subsection{Ethical Concerns}

Since the data provided to us come from actual users
they have to be handled responsibly and with special care.  All data come
anonymized and without any PII in them.  The web browsing logs are further
anonymized by only getting the domain name from the whole URL. Therefore, no
user artifacts (e.g. user names, real names, etc.) are kept in the logs.
Furthermore, all the data analyzed and reported was in aggregated form with no
individual user data accessed or available.
The legal department, of the company
that shared the data, reviewed the paper before submission to ensure that the
data was processed ethically and preserved the customers' anonymity. 

\section{Web-tracking footprint} \label{sec:results} 

In this section, we analyze the output of the crawler and provide estimations
for web-tracking footprint at different granularities: per geographical
location, per tracking companies, per website categories, and per device types.
Thanks to the browsing history reconstruction methodology explained in the
previous section, we also estimate the \COtwo emission rates per an average
user only due to the trackers that exist on the websites browsed.

\subsection{The overall picture}
In Table~\ref{tab:results-countries}, we provide a geographical
breakdown of the users in our dataset. Our telemetry spans over 6
continents and 211 different countries.
While it has great geographic visibility, some
continents (e.g., Africa) or countries (e.g., China) may be under-represented
due to the tech-company's customers distribution. Nevertheless, all the
results provided hereinafter account for this inequality, and they are normalized
considering the \textit{average user} of each continent and its countries. 
The majority of connections (24\%) originate from
North America, and the most frequent countries that appear in our telemetry are 
the United States and Australia, that respectively account for 11.7\% and 10.3\% of the users.
The median user is active during 20 over 28 days of the telemetry collection
period, browses on average 5 out of 7 days per week with a frequency of 4 hours
per day. By looking at aggregated browsing behaviors, we observe that users
visit 197 different domains throughout the 4 weeks, by accessing on average 64
distinct domains per week. 

\begin{figure} 
	\centering
	\includegraphics[width=1\columnwidth, height=0.3\textheight]{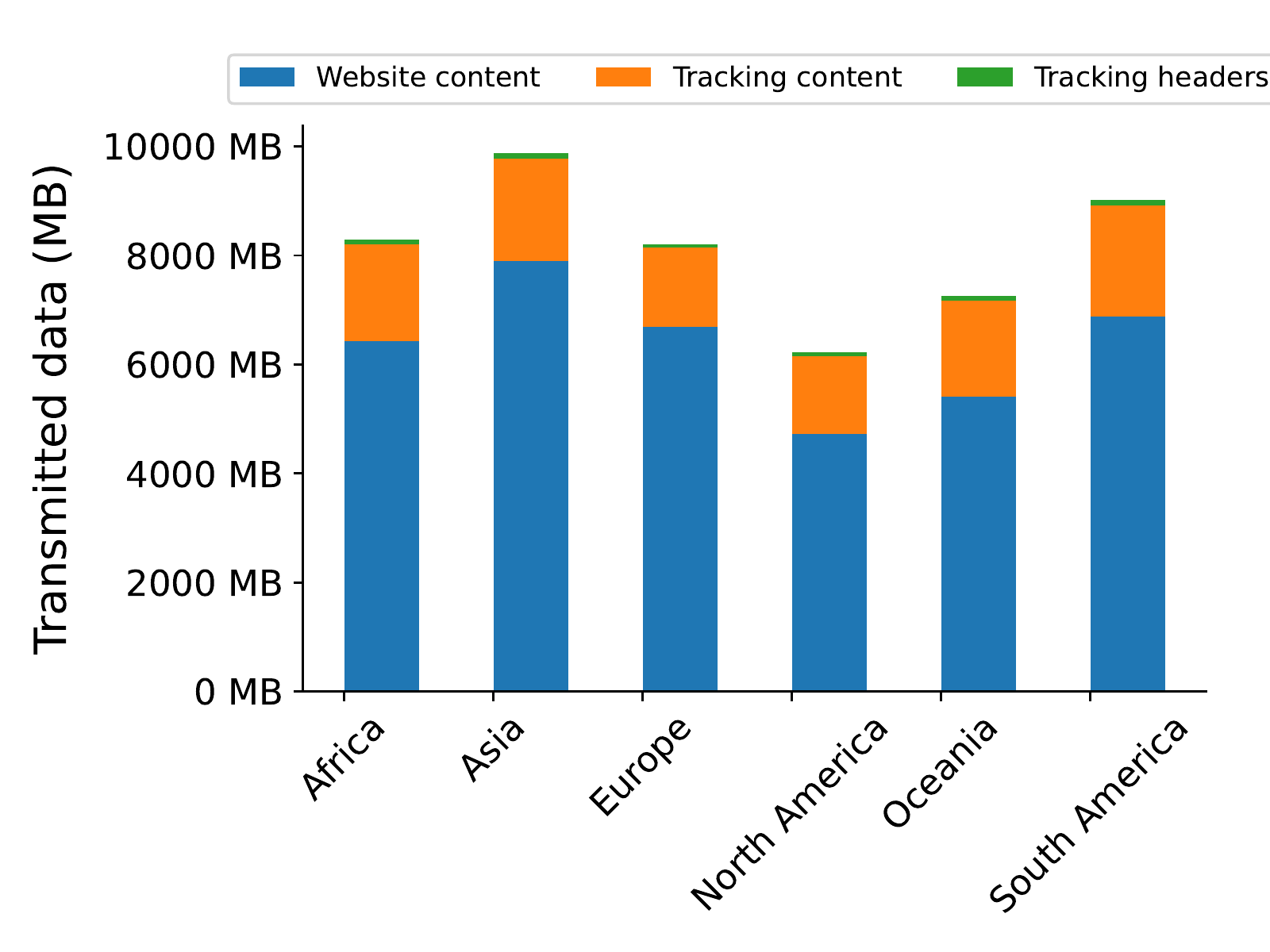}
	\caption{Breakdown of transferred data for the average user in each continent}
	\label{fig:breakdownData} 
\end{figure}

In Figure~\ref{fig:breakdownData}, we report the
breakdown of the yearly transferred data for the average user in each
continent. As our telemetry covers 4 weeks evenly distributed across the year,
we compute the mean amount of data transmitted for each user during one week by
averaging the four weeks of activity at our disposal and project it to one
year.  On average, the annual data downloaded and uploaded by an
\textit{average user} worldwide accounts for 8.14 GB, out of which 1.67 GB are
attributable to tracker-related exchanges (21.11\%).
The aformentioned data do not include multimedia, such as video streams or images, and only refer to data related to content and headers.
Parssinen et. al~\cite{parssinen2018environmental} estimated the 25 to 75\% of the web
contents to be due to online advertisement with high uncertainty. While our
focus is broader than online advertisement, we find the ratio to be lower than
what was known. Note that our work focuses only on the web content excluding
the analysis of other media contents such as images and videos (e.g.  videos
streamed over the network). If media transmitted by trackers were
also take into account, higher numbers would be obtained.

If we project the average data transmitted per capita to the current
active Internet population (i.e., 4.95 B users), we report that web tracking is
responsible for the transmission of 8,415 PB yearly. By further considering 3.1
kWh (1.8 kWh) to transfer 1 GB of data, and an average of 420 g \COtwo emitted per kWh,
we compute that the worldwide emissions due to tracking reaches 10.79 Mt (6.28
Mt) of \GHG.

\subsection{Geographical analysis}

In our telemetry, the highest amount of
transmissions due to web browsing is produced by Asian users, who exchange 9.88 GB per capita
in one year. On the other hand, we observe North American users, who generate on 
average 6.22 GB of network traffic. We found that an average Asian user visits many more
websites than an average North American one (503.76 vs 343.70) during the
period of 4 weeks. Moreover, the mean website visited in Asia is slightly bigger than
the ones browsed in North America (3.83 MB vs 3.69 MB). This explains the
obvious difference observed in Figure~\ref{fig:breakdownData}.
If we look at the biggest data generators due to trackers, we find South America
on top of this list. South-American users generate 2.06 GB of trackers traffic,
Asian 1.89 GB, and African 1.77 GB per year. On the contrary, we observe that
tracking organizations generate less traffic in Europe and North America, 1.45
GB and 1.42 GB respectively.
We have to note that higher data transmissions
do not necessarily mean higher \COtwo emissions.  Indeed the average North
American user will produce 1.42 GB due to tracking whereas the average South
American 2.06 GB.  When taking into account the actual \COtwo emissions we see
that the average US citizen produces 0.45 kg of CO2 per kWh (0.64kg of \COtwo
due to tracking), whereas the average Brazilian only 0.07 kg of CO2 per kWh
(0.14 kg of \COtwo due to tracking).  This is almost 4.6 times more than the
average US citizen compared to the average Brazilian.  We also find that users
from Asian countries such as Japan, Hong Kong, and India produce as much as
0.95, 1.53, and 1.32 kg of \COtwo respectively mostly because of the resources
used for electricity production in those countries.

\begin{figure*}[h]
	\begin{subfigure}{.32\textwidth}
		\includegraphics[width=1\columnwidth]{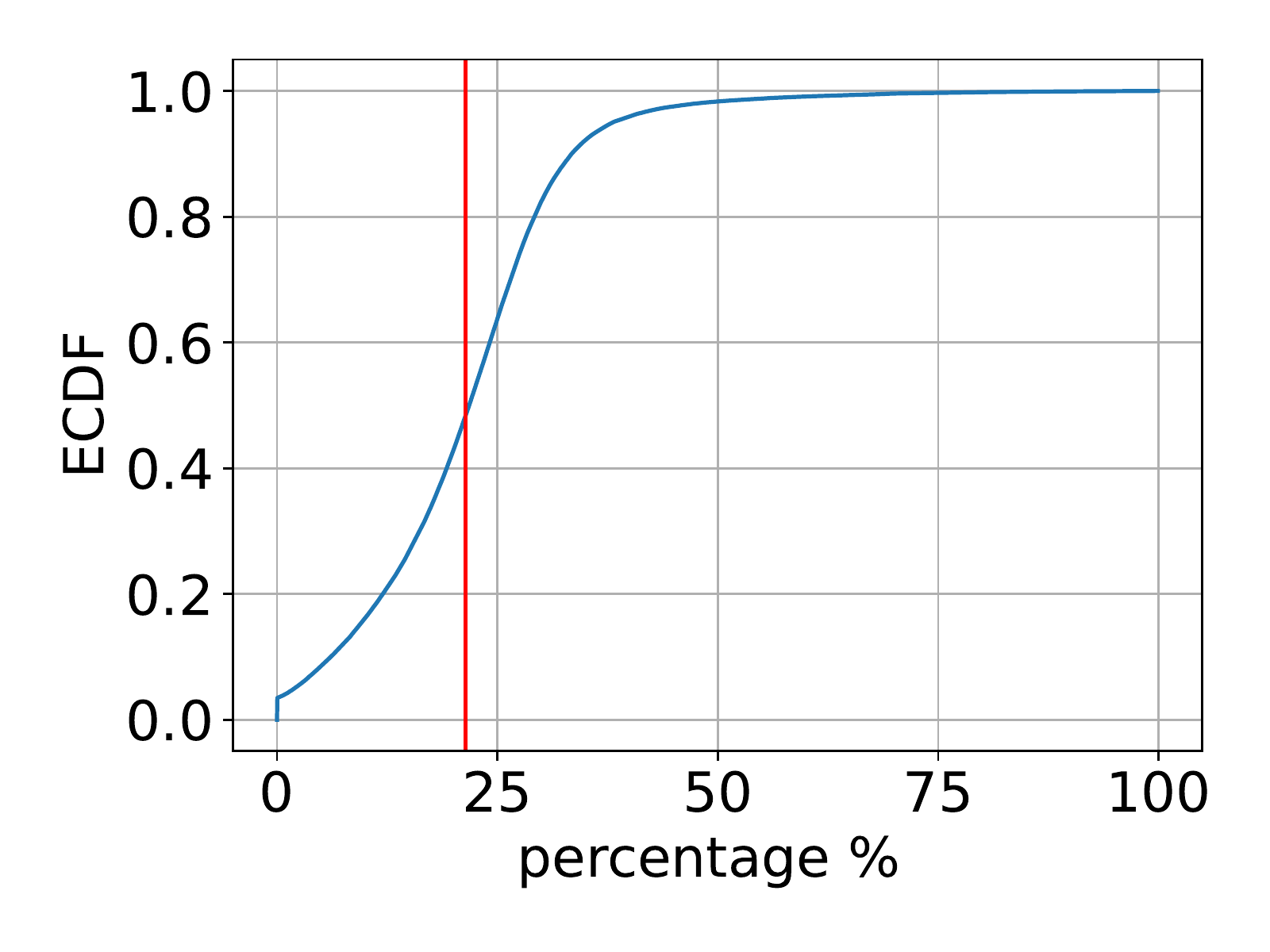}
		\caption{Ratio trackers/webpage}
		\label{fig:ratioTrackersContentECDF}
	\end{subfigure}
	\hfill
	\begin{subfigure}{.32\textwidth}
		\includegraphics[width=1\columnwidth]{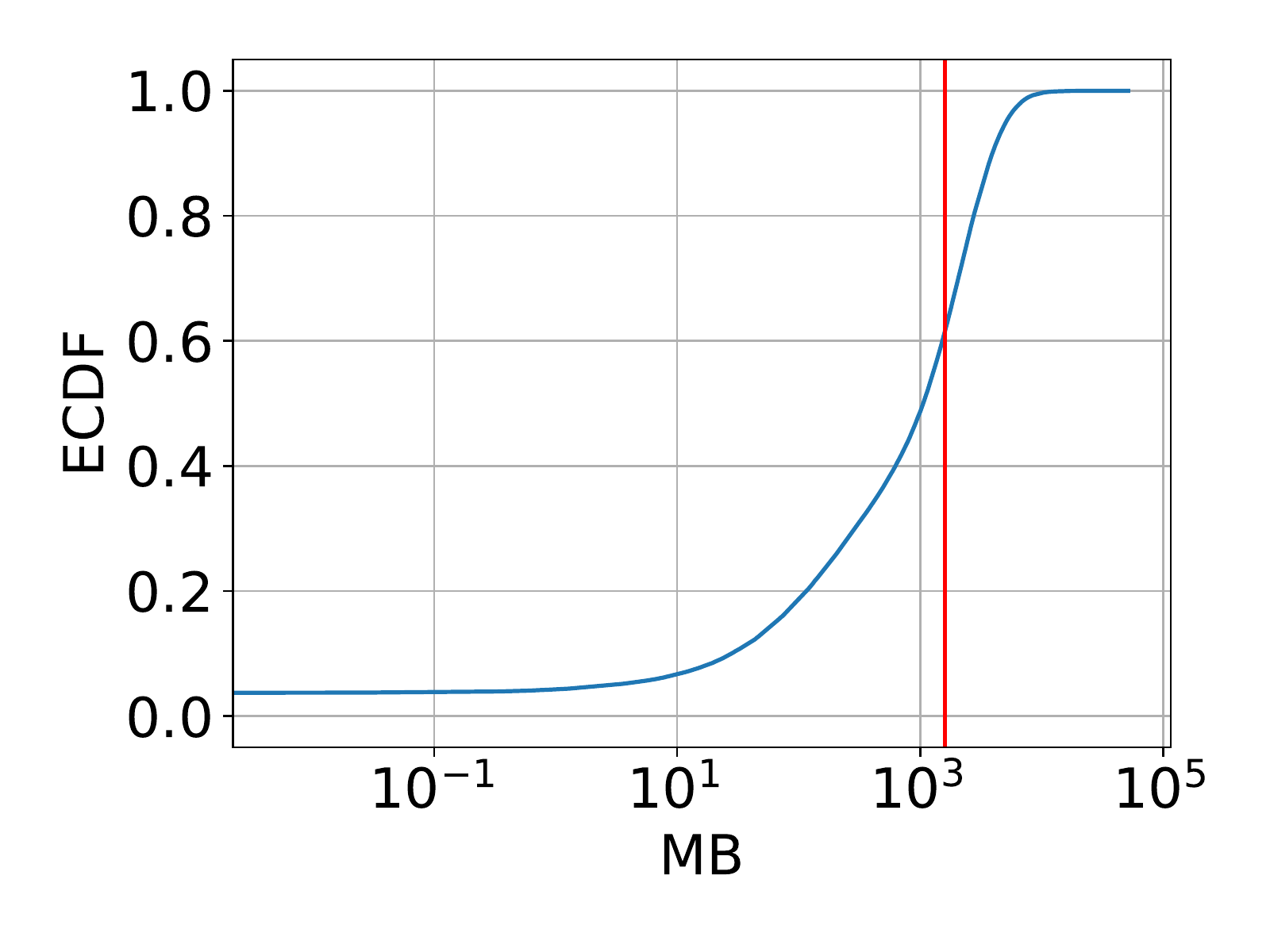}
		\caption{Tracking contents}
		\label{fig:ratioTrackersContentBytes}
	\end{subfigure}
	\hfill
	\begin{subfigure}{.32\textwidth}
		\includegraphics[width=1\columnwidth]{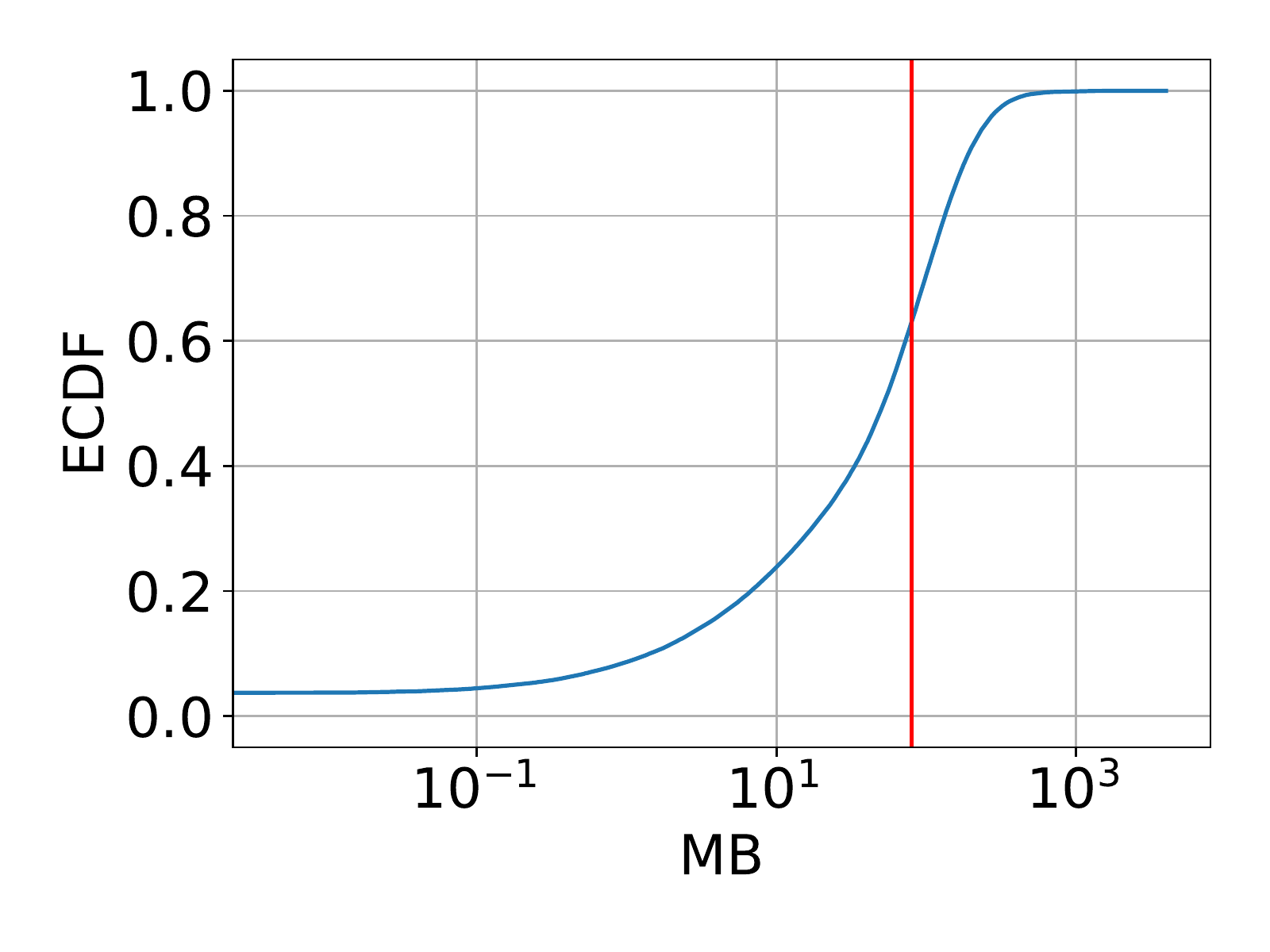}
		\caption{Tracking headers}
		\label{fig:headersTrackersContentBytes}
	\end{subfigure}
	\caption{\textit{Left} - ECDF of ratio between the data transmissions of trackers versus
		the whole website, \textit{Center} - Megabytes transferred due to tracker-related
		traffic, \textit{Right} - Megabytes transferred due to HTTP headers in tracker-related
		traffic}
	\label{fig:TrackersContentECDF}
\end{figure*}

\subsection{User browsing analysis}
In Figure~\ref{fig:TrackersContentECDF},
we report the distribution of ratios between the tracking and the website
contents (left), the megabytes exchanged due to web-tracking
(center), and the traffic generated due to the transmission of the sole
tracker-related HTTP headers. The three figures are complemented with the
corresponding global means. More than 96\% of the active population in our
dataset visits websites whose tracking content is lower 
than 40\% compared to the rest of the website's content.
Similarly, 90\% of the users receive less than 4GB of tracking content annually
and transfer 192 MB of headers related to trackers.
We observed some websites (45,670) whose tracking content represents a considerably
higher percentage (90\%) of the whole data transmitted. 
A deeper analysis revealed that those websites
mostly depend on third-party services or CDNs for content generation, which also
acts as trackers when providing the resources.

\subsection{Categorical analysis}
In Table~\ref{tab:results-categories}, 
we summarize the top-10 categories according to their prevalence
and report the impact that web tracking has on each of them in terms of transferred
data. 
We categorize the websites based on a categorization service provided to us by the AV company.
The service supports over 60 languages and is composed of several specialized modules that disassemble web pages and analyze their
components, such as the webpage language, source code language, documenttype, character set, externallink categories, content words,
scripts and iframes. In addition, the categorization is fine-tuned
by an offline system, which simultaneously analyzes multiple pages looking for connections and 
additional evidence to supplement what was collected in real time. HTTP referrer headers
and hyperlinks are examples of attributes used in this phase.

Generally, we observe a correlation between the most browsed categories
and those that present the highest amount of generated traffic due to trackers.
\emph{Technology, Business,} and \emph{Shopping} websites deviate from the mean
by presenting a higher share of tracking data transmitted yearly per capita ($>$
126 MB). In webpages that serve \emph{News and media} content, we instead measure 
the highest mean overload introduced by tracking resources (i.e., 2.16 MB per
website). When we have a closer look at the categories generating the highest
tracking traffic, we observe that not only the trackers in such websites
include heavier content, but also that they are responsible for a considerably higher
number of web requests. While their mean number is
40.9 in our dataset, an average \emph{Shopping} website requests 
47 resources from third-party trackers, and an \emph{Entertainment}
page 58. Once again, \emph{News/Media} websites are the ones characterized by
the highest mean value of 85 resources related to trackers. 
When examining the average size of tracking-related resources (e.g. javascript
libraries, application/json, etc.), we register a mean size of 70 KB per file
with negligible variability across the categories. 
We also calculated the amount of \COtwo emissions due to tracking that is produced by each category globally based on user visits.
We find Shopping websites to be responsible for the highest emissions (0.82 Mt \COtwo), followed by Technology and Internet (0.72Mt \COtwo) and News/Media
websites (0.70 Mt \COtwo).
Even if Shopping and Technology/Internet are not the most highly tracked categories, they are in the top three of most the visited ones which in turn makes them the biggest contributors of \COtwo out there. 
In contrast, News/Media is much less visited but due to the huge amount of tracking per website, it completes the triplet of the biggest \COtwo emitters.

\begin{table}[t]
	\setlength{\tabcolsep}{2pt}
	\centering
	\begin{small}
		\caption{Web-tracking network footprint for the top-10 categories in our
			dataset.}
		\label{tab:results-categories}
		\begin{tabular}{lrrrrr}
			\toprule
			\multirow{2}{*}{\textbf{Category}} 	&  \multicolumn{1}{c}{\textbf{Unique}}
			&\multicolumn{1}{c}{\textbf{\%}} & \multicolumn{1}{c}{\textbf{Total}}
			& \multicolumn{2}{c}{\textbf{Avg tracking bytes}} \\
			&	\multicolumn{1}{c}{\textbf{users}}
			&\multicolumn{1}{c}{\textbf{users}} &
			\multicolumn{1}{c}{\textbf{tracking}}  	&
			\multicolumn{1}{c}{\textbf{per user}} &
			\multicolumn{1}{c}{\textbf{per website}} \\
			\midrule
			Technology/Internet 		&  89.29 k &  92.12 \%		&	549.83 PB 	&129.64 MB  &930.57 kB \\
			Business/Economy			&  86.84 k &  89.59 \%		&	519.84 PB	&	126.70 MB  &1,115.38 kB \\
			Shopping  					&  73.28 k &  75.60 \%		&	628.13 PB	&	181.67 MB  &1,501.51 kB \\
			Entertainment  				&  66.36 k &  68.46 \%		&	212.54 PB 	& 67.80 MB  &1,788.50 kB \\
			News/Media  				&  65.94 k &  68.03 \%		&	541.20 PB 	&172.64 MB  &2,160.26 kB \\
			Health  					&  65.44 k &  67.51 \%		&	267.19 PB	& 85.85 MB  &1,438.99 kB \\
			Financial Services  		&  64.99 k &  67.05 \%		&	184.78 PB 	& 59.78 MB  &870.29 kB \\
			Education  					&  61.44 k &  63.30 \%		&	172.45 PB	& 59.35 MB  &1,222.33 kB \\
			Travel  					&  57.87 k &  59.70 \%		&	253.48 PB 	& 92.10 MB  &1,347.34 kB \\
			Government/Legal  			&  54.39 k &  56.11 \%		&	77.94 PB 	& 30.13 MB  &844.99 kB \\
			\bottomrule
		\end{tabular}
	\end{small}
\end{table}

\subsection{Tracker analysis}
We now turn the camera on tracking companies 
and estimate the total data transmitted due to their existence in the web ecosystem. 
Table~\ref{tab:results-trackers} summarizes the average number of megabytes transmitted per website and per user related to the
top-10 tracking organizations sorted by their prevalence in our dataset. As
already proven by previous studies~\cite{whenSally,sanchez2021journey}, 
\emph{Google} leads the ranking
not only by the number of users tracked but also for the average number of
megabytes exchanged in one year with a single user (709.31 MB) and for the amount
of information transmitted in each request and response that transfer its
resources (706.98 kB). We also highlight that \emph{Facebook}, the second big
player in our dataset, moves 3 times less data than \emph{Google} for the
average user (198.86 MB) but shows a slightly lower transmission rate for each
website that it tracks (399.66 kB).  \emph{Twitter} represents an interesting
case: while exchanging an annual amount of data with a single user that is
around 12 times smaller than \emph{Google}, the requests of its
tracking-related contents are characterized by a large amount of transmitted
data
(448.19 kB). In this respect, we found that \emph{Google} and \emph{Twitter}
respectively send on average 19 and 17 tracking related resources per tracked
website, which is more than twice the average of 7.72 computed for all the
organizations.
In addition, we measured that the tracking libraries of \emph{Twitter} are smaller 
in size compared to the ones of \emph{Google} ($\sim$60kB vs $\sim$77kB)
and that the former organization often transmits media content such as images
and video that we exclude from our measurements ---although considering the
headers transmitted since those are used for tracking purposes. Moreover, users 
encounter \emph{Twitter} much less compared to \emph{Google} due to the
lower presence of the former across the different websites. 
In turn, the combination of those factors results in high average tracking bytes 
per website exchanged by \emph{Twitter}, but a lower amount of annual tracking
data per user. Other interesting cases are the ones of \emph{Pubmatic} and \emph{Rubicon}, two
of the biggest players in the real-time-bidding advertising~\cite{pachilakis2021rtb} and
header-bidding ecosystems~\cite{pachilakis2019hb}. Those organizations primarily focus on
tracking users and delivering advertisements and don't offer any content or
functionality as other trackers might do. Indeed, when examining the average number of resources
per website, we found them to be on average 4.7 per website for \emph{Pubmatic}, and 5.1
for \emph{Rubicon} ---lower than the overall average of 7.71---, confirming their
main interest only in tracking pixels, auction data and online advertising
libraries.

When we look at the environmental impact in terms of \COtwo emissions, we
estimated that \emph{Google} annually produces more than all the others top trackers combined.
With a whopping 4,382 kt of \COtwo emissions, \emph{Google} dominates the
tracking scene and overshadows other big players such as \emph{Facebook} (1,129 kt), 
\emph{Amazon} (140 kt), and \emph{Microsoft} (98 kt).

\begin{table*}
	\centering
	\caption{Network and carbon footprint of the top-10 tracking organizations along
		with the PB transferred annually due to tracking, and the lower and upper \COtwo emissions.}
	\label{tab:results-trackers}
	\begin{tabular}{lrrrrrrr}
		\toprule
		\multirow{2}{*}{\textbf{Tracker}} &  \multicolumn{1}{c}{\textbf{Unique}} &
		\multicolumn{1}{c}{\textbf{\%}}  & \multicolumn{2}{c}{\textbf{Avg
				tracking bytes}}& \multicolumn{1}{c}{\textbf{Transferred}}&
		\multicolumn{2}{c}{\textbf{\COtwo emissions}}\\ 
		&\multicolumn{1}{c}{\textbf{users}} & \multicolumn{1}{c}{\textbf{users}}  & \multicolumn{1}{c}{\textbf{per
				user}} & \multicolumn{1}{c}{\textbf{per website}} &
		\multicolumn{1}{c}{\textbf{worldwide}} &
		\multicolumn{1}{c}{\textbf{1.8kWh/GB}}
		& \multicolumn{1}{c}{\textbf{3.1kWh/GB}}\\
		\midrule
		Google 			&  92.91 k&  95.86 \% 	&709.31 MB	&706.98 kB	& 3,365.61 PB	&2,544.40 kt&4,382.02 kt\\
		Facebook 		&  85.43 k&  88.14 \% 	&198.86 MB	&399.66 kB	& 867.61 PB 		&  655.91 kt&1,129.63 kt\\
		Adobe			&  75.95 k&  78.36 \% 	& 36.87 MB	&343.35 kB	& 143.01 PB 		&  108.11 kt&  186.20 kt \\
		Microsoft 		&  74.33 k&  76.69 \% 	& 19.75 MB	&136.20 kB	& 74.971 PB 		&   56.68 kt&   97.61 kt  \\
		Amazon 			&  73.78 k&  76.12 \% 	& 28.74 MB	&131.85 kB	& 108.29 PB 		&   81.87 kt&  140.99 kt \\
		Twitter 		&  71.52 k&  73.79 \% 	& 57.99 MB	&448.19 kB	& 211.81 PB 		&  160.13 kt&  275.78 kt \\
		Hotjar 			&  68.17 k&  70.33 \% 	& 32.49 MB	&220.22 kB	& 113.11 PB 		&   85.51 kt&  147.27 kt \\
		PubMatic 		&  67.29 k&  69.42 \% 	&  6.14 MB	&66.40 kB	& 21.10 PB  		&   15.95 kt&   27.47 kt  \\
		Criteo 			&  64.69 k&  66.74 \% 	& 13.92 MB	&102.47 kB	& 45.99 PB  		&   34.77 kt&   59.88 kt  \\
		Rubicon 		&  67.53 k&  69.67 \% 	&  5.22 MB	&23.65 kB\	& 18.00 PB  		&   13.61 kt&   23.44 kt  \\
		\bottomrule                   
	\end{tabular}
\end{table*}

\section{Web-tracking carbon footprint} \label{sec:comparisons}
We estimated that the tracking ecosystem's contribution to global pollution can
be up to 10.76 Mt of \COtwo annually. We now compare web-tracking emissions to
the ones of other human activities, namely those discussed in
Section~\ref{sec:carbonfp}. The comparisons take into account the lax estimation
of 3.1 kWh/GB to convert transferred data to electricity consumed, and the mean
value of420 g \COtwo per kWh to consequently assess the \GHG emitted.

\subsection{Foodprint vs TrackPrint} 
The food industry accounts for $25\%$ of all the
yearly emissions of \COtwo in the atmosphere. Meat production is estimated to be
the biggest offender when considering all of the energy consumed for feeding
the cattle, processing and distributing the meat, and the 
\GHG produced from the birth of the animal to
the end of its life. Taking into account an average cattle weight of 720 kg, 
$40\%$ of consumable meat~\cite{cowPerc}, and using the aforementioned
beef-related emission statistics, we estimate that the emission caused by the
third-party trackers equals the production life-cycle of $416,281$
cows. On the other hand, if we made similar estimates for pigs (i.e., the average
weight of 130 kg, and $70\%$ of consumable meat~\cite{pigPerc}), we would obtain
$5,155,279$ pigs. This translates to 91 and 69\% of the pigs and cows
respectively bred in Argentina and Greece~\cite{pigPerc,cowPerc}. 
Clearly, other types of food such
as dairy products, vegetables, legumes, and nuts, generate megatonnes of \COtwo
every year. Tracking generates \COtwo emissions equivalent to 28\% of the cheese consumption in
France~\cite{cheeseFrance}, $257\%$ of coffee drunk in Italy~\cite{coffeeItaly}, 
and $310\%$ of olive oil produced in Spain~\cite{oilSpain}.

\subsection{Transportation}
Worldwide \GHG imputable to web-tracking practices are
comparable to the number of emissions due to plane and car transportation in the
developed countries listed in Table~\ref{tab:co2_consumptions_and_emissions}. On
average, the impact of web tracking on global pollution is similar to the impact
of 80.61\% of the whole aviation transportation system for the top-10 emitters
in Table~\ref{tab:co2_consumptions_and_emissions} (46.87\% according to our
conservative estimation). At the two extremes, the emission rate of tracking is 4.19
times higher than the whole aviation emissions in Iran (2.56 Mt of \COtwo per
year) and represents 5.57\% of those produced by domestic and international
flights of all the US population in one year. Furthermore, tracking impact is
comparable to 61.32\% and 56.36\% of emissions of more moderate countries such as
Japan and Canada in the context of plane transportation. The environmental
impact of web tracking is also quantifiable by looking at the number of flights
required to emit the same quantity of \GHG.  By considering the figures reported
in section~\ref{sec:carbonfp}, we estimated that a Boeing
737-400 traveling from London to New York with a load factor of 65\% (i.e., with
164 passengers)~\cite{boingConsumption} produces 161.7 tons of \COtwo gases: to
match worldwide web-tracking-generated emissions, the same plane will have to
fly the same route 66,788 times --- which will require 11 thousand years at the
current frequency of 30 flights per day~\cite{londonNewYork}.

Passenger car transportation has been found to contribute five times more than
aviation to global \GHG production
(Section~\ref{sec:carbonfp}). In this context, if we consider
per capita emissions of the US ---the country with the highest level of \COtwo
emissions due to passenger cars--- the overall environmental impact of data
transmissions related to tracking corresponds to the annual amount of \COtwo
produced by the cars of 2.4M US citizens. In India and China, the same amount
corresponds to the yearly \COtwo production of 54M and 21.6M inhabitants' cars.
To provide the reader with a concrete example that allows us to better quantify
web-tracking carbon footprint, we consider the \COtwo produced by an average car
(i.e., 122.3 g\COtwo/km) during the trip from Rome to Paris (i.e., 1412.6km). To
match the annual carbon emissions of tracking for the global Internet
population, the car will have to travel from one city to the other 62.12M times
and accumulate 87.75B kilometers. From a different perspective, this would
correspond to going around the world along the equator 2.2M times.

\subsection{Cryptocurrency}
Bitcoin (BTC) is the most important cryptocurrency by
market cap (over $\$850B$). 
It is based on the proof-of-work consensus model, in
which heavy computational effort is needed in order to validate transactions in
the blockchain. 
As we discussed in Section~\ref{sec:method}, Bitcoin is annually 
responsible for over 70 megatonnes of \COtwo emissions, with the US and Russia
being the biggest contributors. 
Our estimations for the web tracking ecosystem correspond to $72\%$ of all the
bitcoin-related emissions in the US, and $121\%$ of the ones in
Russia. If we calculated some global transaction-level statistics~\cite{transCrypto}, we
would find that tracking emissions are equal to $24,087,768$ bitcoin transactions
(almost 2 full months of the global bitcoin transactions). Taking into
consideration the median Bitcoin transaction value of 0.013 BTC~\cite{valCrypto},
and the current BTC price of $\$43,455$~\cite{dollarCrypto}, it would be equal
to $\$13B$.

\subsection{Electricity of Home Appliances} 
The residential electricity
consumption is estimated to be responsible for 20\% of the energy-related \COtwo
emissions~\cite{homeappliances}, which is by the far the biggest contributor to
the overall pollution. Among the home appliances, heaters and coolers
are identified to consume the majority of the household electricity. An
average household, where heaters are run for 7.5 hours per day during the
average heating season of 5.6 months~\cite{heatingavg}, produces 6.3 tonnes of
\COtwo every year. If third-party tracking did not exist or all trackers
were blocked, 1.6M households could be heated while emitting the same quantity of
\COtwo.
From another perspective, the electricity required by the tracking ecosystem
would heat half of all the houses in New York City for one full year~\cite{newYorkHeat}.

\section{Related Work} \label{sec:related} As discussed in the introduction of
this work, an estimation of the impact that Internet transmissions have on the
environment is complex to achieve, mainly because it requires computing how
much electricity is consumed to transfer the data and to quantify the amount of
\COtwo emitted to produce it.  Both aspects have been explored by scattered
works, with very different results and high uncertainty.

\textbf{Energy consumption of computer networks} - In their
work~\cite{baliga2009energy}, Baliga et al. modeled the internet infrastructurei
and estimated from users' usage an energy consumption of 0.04 kWh per
transmitted gigabyte.  With a similar approach, Costenaro et
al.~\cite{costenaro2012megawatts} found the same estimation to be 5.12 kWh/GB.
In their works, Schiem et al, proposed a model that takes into account users'
devices and calculated that a gigabyte of transmitted data implies the
consumption of 0.052 kWh~\cite{coroama2015energy,schien2015energy}.
Coroama et al. directly measured the energy consumption of transmitting data
through the network by using real-world data and found an overall cost of 0.1993
kWh / GB~\cite{coroama2013direct}.

Finally, Aslan et al. leveraged previous studies on electricity consumption due
to Internet transmissions, and measured that a single gigabyte requires 0.05 kWh
to be transmitted~\cite{aslan2018electricity}.

\textbf{Web-tracking \COtwo emissions} - In their report, Cucchietti et al.
crawled the top-1M websites from the Tranco list and estimated the carbon
footprint of web-tracking by only collecting the cookie names and their
values~\cite{cucchietti2022tracking}. Authors based their analysis on several
assumptions about the generated traffic on those websites, and the number of
cookies they serve. Their methodology allowed to compute that tracking-related
traffic due to cookie transmissions is globally responsible for the emissions of
11 Mt \COtwo each year.
On the other hand, P{\"a}rssinen et al. investigated the emissions of online
advertisement, with a particular focus on fraudulent
practices~\cite{parssinen2018environmental}.  By aggregating previous studies,
the authors proposed a framework to estimate the environmental impact of web
advertising on both desktop and mobile ecosystems.  Due to the high uncertainty
in those works, their estimation varies from 12 to 159 Mt \COtwo.
Uijttewaal et al. focused instead on the mobile ecosystem and studied how
tracking and advertising in smartphone applications contribute to \GHG
emissions. The authors used data from previous studies and roughly estimated that
the total emissions could range from 5 to 14 Mt \COtwo per year in Europe, with
unsolicited tracking generating from 3 to 8 Mt \COtwo.

\section{Limitations}
\label{sec:limitations}
Our work tries to estimate the CO2 emissions of online tracking.
Our main focus is to understand and calculate the amount of data that third
party companies transmit to track users by analyzing real user browsing
behavior.  Computing the energy consumption of computer networks and equipment
is out of the scope of this work and for this reason we have to rely on previous
energy calculations.  Even so, we used the most recent and widely accepted
calculations to reflect as good as possible the actual CO2 emissions of online
tracking on the internet at the current moment.

Our study serves as an estimation of online tracking emissions, and
not a general projection at any point in time, since online tracking is an
ever-evolving ecosystem that constantly changes by introducing new technologies
and mechanism. The approach here presented can serve as guideline for future
studies on the topic.  

Due to situations of closed gardens or paid subscriptions, we are not
able to login into certain services. This means that we may collect a lower
bound in some situations, since more tracking might occur when a logged user
browses through the website.

Finally, the browsing history of the users consists of mobile and
desktop users.  Mobile and desktop devices consume different amounts of
electricity due to different specs (e.g. network connectivity, CPU consumption,
screen energy usage, etc.), so the energy consumption could be different between
those devices.  In this study we use generally accepted energy consumption
numbers in order to transfer 1 GB of data independently of the user device.  The
differentiation between the specific consumption in mobile and desktop users is
out of the scope of this study and is left as future work.

\section{Discussion} \label{sec:discussion} 
In this study, we try to estimate the \COtwo emissions produced by the
web-tracking ecosystem and found those to account for almost 11 Mt of \COtwo annually. We report
as well that this value is comparable to many activities of
modern life, such as meat production and consumption, transportation, and even
cryptocurrency mining. Considering that a tree can on average absorbs 20 kg of
\COtwo every year, we would need to plant 550 M of them to compensate for the
emissions due to web tracking.  Our analysis is to our knowledge the most
comprehensive study so far which put the tracking ecosystem under the spotlight,
investigating an understudied but critically-important aspect of modern
society. As broadly discussed, such estimation is not trivial to obtain:
although we consider real-user telemetry, diversify their geographical
locations, and provide an upper and a lower estimation, we highlight that our
results represent a lower bound.  The tracking ecosystem is far more complex and
opaque than what is visible to the users. Indeed, there is a plethora of
connections and transferred data behind the scenes.  Trackers and advertisers
continuously exchange information with each other and by using their internal
networks: a measure of these data transfers from the user's perspective is
impossible to achieve since the whole process is completely opaque.  In any
case, the existence of such networks and interconnections leads to much higher
\COtwo emissions, which adds to the one that we could estimate from our
perspective.

Is there something that users could do to limit the emissions due to web
tracking? The answer to this question is multilayered. Firstly, users could
significantly reduce the information that trackers have by utilizing
anti-tracking solutions.  This approach would not only reduce data
transmissions but also decrease the computation resources needed to process a
web page, in turn saving electricity. As a consequence, tracking organizations
would have fewer data to share, which would further limit polluting emissions.  A
more drastic approach would be the installation of specific software at the
network level (i.e., in the routers) to completely drop connections towards
trackers. This can further reduce \COtwo emissions as the data never arrives at
the users' devices, and does not need to be processed. Such possibility is however
too strict, as in many cases it would prevent some functionalities and 
would disrupt the whole tracking and advertising ecosystem.

It is interesting to note that our study differs significantly from previous studies.
Our methodology accounts for the all the possible techniques and tracking
activities from third-party trackers,
by monitoring them in a network level (e.g. finding requests being sent and
received, monitoring scripts third-party trackers load, etc.)

In contrast to previous works, we don't focus on specific tracking techniques such as cookies, which will limit our visibility into the tracking ecosystem.
Also since not all users visit the same websites and not all websites have the same number of cookies, tracking scripts and tracking techniques it is not safe to assume just averages.
In this study, we use a systematic and thorough approach that considers multiple different factors.
Our approach is designed to represent the whole tracking ecosystem, and not just some aspects of it (e.g. cookies).
In this work, we focus on real user behavior, in contrast to previous studies,
so our calculations rely as little as possible to estimations and projections.
To the best of our knowledge, this is the most reliable way of representing
third-party tracking \COtwo emissions to date.
Finally, our methodology can be easily generalized since it does not rely on estimations on user behavior or global averages.

\section{Conclusion} 
\label{sec:conclusions} 

In this work we try to estimate the \COtwo emissions of the tracking ecosystem annually across the globe.
By designing a thorough analysis that takes into account multiple factors regarding online tracking we were able to calculate the \COtwo emissions due to tracking activities in a systematic and comprehensive manner.
We utilize the browsing telemetry of 100K users to design a comprehensive experiment across 41 million website visits (2.7 million unique domains).
We observe that web tracking increases data transimission up to 21\% annually per user which further increases \GHG  emissions.
We found that the total emissions are up to 10.76 Mt of \COtwo which is comparable to many aspects of the modern life such as meat consumption, aviation and transportation and even cryptocurrency mining.
We conduct a multilayered study broken down by user continents, country energy production mixture and even resource caching across websites.
Our analysis is to our knowledge the most comprehensive study so far which put the tracking ecosystem under the spotlight and reports an understudied aspect but critically important for the modern society.

\bibliographystyle{ACM-Reference-Format}
\bibliography{biblio}

\end{document}